\documentclass[11pt]{article} % use larger type; default would be 10pt

\usepackage[utf8]{inputenc} % set input encoding (not needed with XeLaTeX)

\usepackage{geometry} % to change the page dimensions
\usepackage[dvips]{graphicx}
\DeclareGraphicsExtensions{.bmp,.png,.pdf,.jpg}
\usepackage{booktabs} % for much better looking tables
\usepackage{array} % for better arrays (eg matrices) in maths
\usepackage{paralist} % very flexible & customisable lists (eg. enumerate/itemize, etc.)
\usepackage{verbatim} % adds environment for commenting out blocks of text & for better verbatim
\usepackage{subfig} % make it possible to include more than one captioned figure/table in a single float
\usepackage{fancyhdr} % This should be set AFTER setting up the page geometry
\usepackage{sectsty}
\allsectionsfont{\sffamily\mdseries\upshape} % (See the fntguide.pdf for font help)
\usepackage[nottoc,notlof,notlot]{tocbibind} % Put the bibliography in the ToC
\usepackage[titles,subfigure]{tocloft} % Alter the style of the Table of Contents

\author{ A. Sandoval-Villalbazo and A.R. Sagaceta-Mejía
	\bigskip \\
	Department of Physics and Mathematics,  U. Iberoamericana \\ M\'exico City, M\'exico.}

\title{Jeans instability for an inert binary mixture: a kinetic theory approach in the Euler regime}

\begin{document}

\maketitle
\begin{abstract}
\noindent The Jeans stability criterium for gravitational collapse  is examined for the case of an inert binary mixture in local equilibrium, neglectinq dissipative effects. The corresponding transport equations are established using kinetic theory within the Euler regime approximation. It is shown that the corresponding dispertion relation is modified, yielding corrections to the Jeans wave number that can be generalized for several interesting cases involving dissipation.
\end{abstract}

\section{Introduction}
Structure formation through gravitational collapse is one of the most important processes present in astrophysics. First identified by James Jeans in 1902 \cite{Jeans1}, gravity acts through very long scales allowing density fluctuations to grow exponentially when the mass present at certain scales surpasses a given mass thresehold.  A very simple argument that leads to a fairly accurate approximation of this threshold consists in the consideration of the wave equation for the density fluctuations $\delta \rho$, valid for an isothermal one-component system with equilibrium density $ \rho_{0}$ \cite{KT, SV1}:

\begin{equation}
\nabla^2 \delta \rho-\frac{1}{C^{2}_{T}} \frac{\partial^2 \left(\delta \rho\right)}{\partial t^2}+\frac{4 \pi G  \rho_{0}}{C^{2}_{T}} \delta \rho=0,
\label{W1}
\end{equation}

\noindent where $C^{2}_{T}$  is the isothermal speed of sound and $G$ is Newton's gravitational constant. The wave-like solution $ \delta \rho= A \exp [i ( \vec{q} \cdot  \vec{r}- \omega t)] $ leads to the \textit{dispersion relation}:

\begin{equation}
 \frac{\omega^2}{C^{2}_{T}}-q^2+\frac{4 \pi G \rho_{0}}{C^{2}_{T}}=0.
\label{W2}
\end{equation}

 \noindent Growing exponential modes associated to structure formation correspond to the existence of  imaginary roots of Eq. (\ref{W2})

\noindent The corresponding thresehold is given by the \textit{Jeans wavenumber}:
\begin{equation} 
q_{J}=\sqrt{\frac{4 \pi G \rho_{0}}{C^{2}_{T}}}.
\end{equation}

Now, since the wavelenth of the system is related to the wave number through the relation $\lambda = \frac{2 \pi}{k}$, a rough estimation of the critical mass needed in order to form a structure through gravitational collapse in a volume scaled as $\lambda^3$ reads:

\begin{equation}
M_{J} \simeq \frac{\pi^{3/2} C^{3}_{T} }{G^{3/2} \rho^{1/2}_{0}}.
\label{MJ}
\end{equation}

\noindent For the case of a molecular cloud $\rho_{0} \sim 1.6 \times 10^{-19} \frac{kg}{m^3} $ and  $C_{T}\sim 0.435 \times 10^{2} \frac{m}{s} $, yielding $ M_{J} \sim 2.1 \times 10^{30} kg $, which is quite close to the value of a solar mass.  \\

\noindent The calculation that leads to Eq. ($\ref{MJ}$) has been generalized by several authors in order to include dissipative effects such as viscosity and heat conduction  for the case of one component dilute gases \cite{SV1,Weinberg,CG,Carle}. Heat fluxes associated to density gradients (Dufour effect) have been addressed for the case of a dilute two component plasma \cite{SV2}. The case of a binary mixture in the Euler regime has been only  partially addressed in phenomenological approaches and needs to be revised in by means of a kinetic theory formalism \cite{AA, GZ}. \\

\noindent The purpose of the present paper is to fill this gap giving the kinetic foundations of the derivation of the dispersion relation for an inert dilute binary gas in the presence of Newtonian gravity. In order to accomplish this task, the paper has been divided as follows: section two corresponds to the establishment of the evolution equations of the averages of the conserved microscopic  quantities (mass, linear momentum and energy), special emphasis is made in the structure of the mass fluxes for each type of particle.   In section 3 the transport equations are linearized so that a suitable dispersion relation is established identifying the second component in the stability of the system. Finally, section four is devoted to a discussion of the results here obtained. 

\section{Balance equations}
 
\subsection{Evolution of the distribution function}
The starting point of the formalism is the Boltzmann equation for two species in the presence of Newtonian gravity within the BGK approximation: 
\begin{equation}
\frac{\partial f_{[i]}}{\partial t}+v^{a}_{[i]} \frac{\partial f_{[i]}}{\partial x^{a}}+g^{a}_{[i]} \frac{\partial f_{[i]}}{\partial v_{[i]}^{a}}=-\frac{f_{[i]}-f^{(0)}_{[i]}}{\tau},\quad i=1,2.
\label{B1}
\end{equation}
In this work, the notation $[.]$ is used to identify the $i_{th}$-species and $(.)$ is used to denote the order of the distribution function with respect to the Kundsen number. In Eq. (\ref{B1}),   $\tau$ is the relaxation time to the equilibrium functions $f_{[i]}^{(0)} $ and the  Newtonian gravitational fields $g^{a}_{[i]}$ are expressed in terms of the corresponding scalar potential $\varphi_{[i]}$ given by

\begin{equation}
g^{a}_{[i]}= -\frac{\partial \varphi_{[i]}}{\partial x^a},\quad i=1,2.
\end{equation}

%\begin{equation}g^{a}_{[2]}= -\frac{\partial \varphi_{[2]}}{\partial x^a}\end{equation}
\noindent which in turn satisfies the Poisson equation:
\begin{equation}
\nabla^2 \varphi_{[i]}= 4 \pi G \rho_{[i]}, \quad i=1,2.
\end{equation}
The total gravitational potential $\varphi$ satisfies the equation 
\begin{equation}
\nabla^2 \varphi= 4 \pi G \rho= 4 \pi G (\varphi_{[1]}+\varphi_{[2]})
\end{equation}
Macroscopic variables are identified with statistical averages, these are given, for species with number density $n_{[i]}$, by:
\begin{equation}
\left\langle \psi_{[i]} \right\rangle =\frac{1}{n_{[i]}} \int  \psi_{[i]}  f_{[i]} d\vec{v}_{[i]}.
\label{average}
\end{equation}
The center of mass velocity for the binary system is defined as
\begin{equation}
u^{a}=\frac{m_{[1]}n_{[1]} \left\langle v^{a}_{[1]} \right\rangle+m_{[2]}n_{[2]} \left\langle v^{a}_{[2]} \right\rangle }{m_{[1]} n_{[1]}+m_{[2]} n_{[2]}},
\label{CM} 
\end{equation}
and the \textit{chaotic} velocity for each species reads:
\begin{equation}
c^a_{[i]}=v^{a}_{[i]}-u^{a}
\label{cv}
\end{equation}
Eqs. (\ref{CM}-\ref{cv}) lead to the important relation:
\begin{equation} 
m_{[1]} n_{[1]}  \left\langle c^a_{[1]} \right\rangle+m_{[2]} n_{[2]}  \left\langle c^a_{[2]} \right\rangle=0.
\label{eq:cero}
\end{equation} 

\subsection{Mass balance}
The mass balances for each species are obtained multiplying Eqs. (\ref{B1}) by the collision invariants $m_{[1]}$ and $m_{[2]}$  and upon integration in the respective velocity spaces, yielding:
\begin{equation}
\frac{\partial \rho_{[i]}}{\partial t} + \frac{\partial}{\partial x^{a}} \left(\rho_{[i]} v^a_{[i]}\right)=0,
\quad i=1,2.
\label{M1}
\end{equation}

%\begin{equation}\frac{\partial \rho_{[2]}}{\partial t} + \frac{\partial}{\partial x^{a}} \left(\rho_{(2)} u^a_{[2]}\right)=0\label{M2}\end{equation}
 
\noindent In Eq. (\ref{M1}),  $\rho_{[i]}=m_{[i]} n_{[i]} $ and use has been made of the definition of statistical average (Eq. \ref{average}). It is interesting to notice that  dissipative contributions may appear in the mass balance equations through Eq. (\ref{cv}):

\begin{equation}
\frac{\partial \rho_{[i]}}{\partial t} +\frac{\partial}{\partial x^{a}} \left(m_{[i]}n_{[i]} u^a\right)= -\frac{\partial}{\partial x^{a}}J^{a}_{[i]}, \quad i=1,2,
\label{M3}
\end{equation}

%\begin{equation}\frac{\partial \rho_{[2]}}{\partial t} +\frac{\partial}{\partial x^{a}} \left(n_{(2)}u^a\right)=-\frac{\partial}{\partial x^{a}}J^{a}_{[2]}\label{M4}\end{equation}
\noindent where
\begin{equation}
J^{a}_{[i]}=m_{[i]}  n_{[i]}  \left\langle c^a_{[i]} \right\rangle,
\label{M5}
%\label{M6}
\end{equation}
To complete the mass balance, we notice that the addition of Eqs. (\ref{eq:cero})  over both species lead to the mass balance for the mixture:
\begin{equation}
\frac{\partial \rho}{\partial t} +\frac{\partial}{\partial x^{a}} \left(\rho u^a\right)=0,
\label{M44}
\end{equation}
where the total density of the system is:
\begin{equation}
\rho=\rho_{[1]}+\rho_{[2]}.
\end{equation}

\subsection{Momentum balance}
Upon multiplication of  Eqs. (\ref{B1}) by the collision invariants $m_{[1]}  v^a_{[1]} $ and $m_{[2]} v^a_{[2]}$, the addition of the equations over both species leads  to the density momentum balance equation:

\begin{equation}\frac{\partial }{\partial t} \left(\rho u^a\right)+ \frac{\partial}{\partial x^{b}} \left(\rho u^a u^b+  p \delta^{ab}+\Pi^{ab}\right)=\rho\nabla \varphi\label{Mom1}\end{equation}

\noindent where $p$ is the total pressure $p=p_{[1]}+p_{[2]}$ and   $\varphi=\varphi_{[1]}+\varphi_{[2]}$.
  The last equation has been obtained considering Eq.  (\ref{CM}) and the corresponding averages through the distribution functions:

\begin{equation}
f_{[i]}=f_{[i]}^{(0)}+f_{[i]}^{(1)}.
\label{CE1}
\end{equation}
The equilibrium distribitution function $f_{[i]}^{(0)} $ is given by:

\begin{equation}
f_{[i]}^{(0)}=n_{[i]}
\left(\frac{m_{[i]}}{2\pi k T}\right)
^{\frac{3}{2}}e^{-\frac{m_{[i]} c^2_{[i]}}{2 k T}}.
\label{Max1}
\end{equation}
where $k$ is the Boltzmann constant and $T$ is the  temperature of the system. In Eq. (\ref{Mom1}) 
\begin{equation}
\int_{\Gamma} 
f_{[i]}^{(0)} c^a_{[i]} c^b_{[i]} d\vec{c}_{[i]}=n_{[i]} k T \delta^{a b},
\end{equation}
and
\begin{equation}
\Pi^{ab}=\int_{\Gamma} f_{[1]}^{(1)}  c^a_{[1]} c^b_{[1]} d\vec{c}_{[1]}+\int_{\Gamma} f_{[2]}^{(1)}  c^a_{[2]} c^b_{[2]} d\vec{c}_{[2]}.
\end{equation}
where $\Gamma$ corresponds to all the velocity  space; $\Pi^{ab} $ vanishes in the Euler regime.

\subsection{Energy balance}
Multiplying Eqs. (\ref{B1}) by the collision invariants $\frac{1}{2}m_{[1]}  v^2_{[1]} $ and $\frac{1}{2}m_{[2]} v^2_{[2]}$, the addition of both equations leads to the energy density balance equation:

\begin{equation}
\frac{\partial}{\partial t}
\left(\frac{1}{2}\rho u^2+ \frac{3}{2}\rho k T\right)
+ \frac{\partial}{\partial x^{b}} \left(\frac{1}{2}\rho u^2 u^b + \frac{3}{2}\rho k T u^b+ p u_a\delta^{ab}+J^{b}_{[q]}\right)
=0
\label{E1}
\end{equation}
where the \textit{heat flux} $J^{a}_{q}$ is given by:

\begin{equation}
J^{a}_{q}=\frac{1}{2} m_{[1]} n_{[1]} \left\langle c^2_{[1]} c^a_{[1]}\right\rangle+
\frac{1}{2} m_{[2]} n_{[2]} \left\langle c^2_{[2]} c^a_{[2]}\right\rangle.
\label{H1}
\end{equation}
This heat flux vanishes in the Euler regime, but in the Navier-Stokes case $f^{(1)}_{[i]} $ leads to  useful expressions that contain the Fourier law and the  Dufour  effect \cite{SV2}.

\section{Linearized transport equations}
Eqs. (\ref{M1}), (\ref{Mom1}) and (\ref{E1}) constitute a set of non-linear partial differential equations which can be studied close to local equilibrium in terms of linear perturbations. In this type of formalism the vector containing the unknown variables $x$ is expressed as the sum of its equilibrium value $x_0$ and its corresponding fluctuations $\delta x$:
\begin{eqnarray}
x = x_0+\delta x,\\
\left( \rho_{[i]}, \vec{u} ,T ,\varphi \right)= & \left(\rho_{0[i]},\vec{u}_0, T_0, \varphi _0\right)+\left(\delta \rho_{[i]}, \delta \vec{u}, \delta T, \delta \varphi\right).
\end{eqnarray}
Neglecting second order terms, assuming a vanishing center of mass velocity and defining $\delta \theta =\nabla \cdot \delta \vec{u}$, the linearized set of equations is easily obtained by direct substution. The linearized mass balance reads:

\begin{equation}
	\frac{\partial}{\partial t}\left(\delta \rho_{[i]}\right)+\rho_{0[i]}\delta\theta=0,\quad i=1,2.
	\label{L1}
\end{equation}
The longitudinal mode of the linear momentum balance equation becomes:
\begin{equation}
	\frac{\partial}{\partial t}\left(\delta\theta\right)+
	\left( \frac{\rho_{0[1]}}{\rho_{0}} C^2_{T[1]}
	+\frac{\rho_{0[2]}}{\rho_{0}}C^2_{T[2]}
	\right)\nabla^2 \left(
	\frac{\delta T}{T_{0}}
	\right)
	+ C^2_{T[1]} \nabla^2 \left( \frac{\delta \rho_{[1]}}{\rho_{0}} \right)
	+ C^2_{T[2]} \nabla^2 \left( \frac{\delta \rho_{[2]}}{\rho_{0}} \right)=-\nabla^2 \delta \varphi,
	\label{L2}
\end{equation}
and the set is closed using the linearized energy balance equation, together with the Poisson equation, respectively:
\begin{equation}
\frac{\partial}{\partial t}
\left(\delta T\right)
+\frac{2}{3} T_{0}  \delta \theta=0,
\label{L3}
\end{equation}

\begin{equation}
	\nabla^2 (\delta \varphi) = 4 \pi G\left(\delta \rho_{[1]}+\delta \rho_{[2]}\right).
	\label{L4}
\end{equation}

\section{Dispertion equation and Jeans stability criterium}

The linearized transport equations (\ref{L1}-\ref{L4}) can be treated algebraically   taking successively in each equation the Fourier tansform in space and the Laplace transform in time.  \\Defining:

\begin{equation}
\delta \tilde{X}(\vec{q},s)=\int^{\infty}_{0} \int^{\infty}_{-\infty} \delta x(\vec{r},t) e^{i \vec{q} \cdot \vec{r}} e^{-s t} d\vec{r} dt
\end{equation}
the linearized  transport equations can be immediately expressed as:
\begin{equation}
A \tilde{X}(\vec{q},s)=0
\end{equation}
where
\begin{equation}
\delta \tilde{X}=( \delta \tilde{ \rho}_{[1]}, \delta \tilde{ \rho}_{[2]},  \delta \tilde{ \theta},  \delta\tilde{ T}, \delta \tilde{\varphi})
\label{D1}
\end{equation}
and

\begin{center}
	
\[
A=\left(\begin{array}{ccccc}
\frac{s}{\rho_{0[1]}} & 0 & 1 & 0 & 0\\
0 & \frac{s}{\rho_{0[2]}} & 1 & 0 & 0\\
-\frac{C_{T[1]}^{2}}{\rho_{0}} & -\frac{C_{T[2]}^{2}}{\rho_{0}} & s & -\left(\frac{C_{T[1]}^{2}\rho_{0[1]}+C_{T[2]}^{2}\rho_{0[2]}}{T_{0}\rho_{0}}\right) & -q^{2}\\
0 & 0 & \frac{2}{3} & \frac{s}{T_{0}} & 0\\
4\pi G & 4\pi G & 0 & 0 & q^{2}
\end{array}\right).
	\label{D15}
\]

\end{center}

\noindent Real values for $s$ in the \textit{dispertion relation} are identifified with exponentially growing/decaying modes. In the Euler regime, and using the center of mass velocity in the balance equations (Eq. \ref{CM}), a very simple dispertion relation is obtained. The resulting expression reads: 

\begin{equation}
s^2 [s^2-4 \pi G (\rho_{[1]}+\rho_{[2]})+q^2(\frac{5}{3}C^{2}_{T [1]} \frac{ \rho_{[1]}}{ \rho_{[0]}}+\frac{5}{3}C^{2}_{T [2]} \frac{ \rho_{[2]}}{ \rho_{[0]}})]=0
\label{Disp1}
\end{equation}

\noindent The critical wavenumbers correspond to non-trivial real values of $s$ in Eq. (\ref{Disp1}), yielding:
\begin{equation}
q^{2}=\frac{4\pi G\left(\rho_{0[1]}+\rho_{0[2]}\right)}{\frac{5}{3}C_{T[1]}^{2}\frac{\rho_{0[1]}}{\rho_{0}}+\frac{5}{3}C_{T[2]}^{2}\frac{\rho_{0[2]}}{\rho_{0}}}.
\label{NW}
\end{equation}
The  Jeans Mass for the binary system is established following the same arguments included in the introduction. The result obtained is:
\begin{equation}
M_{J}\simeq\frac{\pi^{\frac{3}{2}}\left(\frac{5}{3}C_{T[1]}^{2}\frac{\rho_{0[1]}}{\rho_{0}}+\frac{5}{3}C_{T[2]}^{2}\frac{\rho_{0[2]}}{\rho_{0}}\right)^{3}}{G^{\frac{3}{2}}\left(\rho_{0[1]}+\rho_{0[2]}\right)^{\frac{1}{2}}},
\label{NJM}
\end{equation}
which reduces, taking $ \rho_{0 [2]}=0$ and  $ \rho_{0 [1]}=\rho_{0}$, to the Jeans mass for a single component system Eq. (\ref{MJ}).
\section{Final Remarks}
The use of a representation of the local variables involving the center of mass velocity   allows to establish simple expressions for the dispertion relation (Eq. \ref{Disp1}). Indeed, for a three-component system, the corresponding matrix in the Fourier-Laplace space reads:

%\begin{center}
	
$$B=\left(
	\begin{array}{cccccc}
	\frac{s}{\rho _{0[1]}} & 0 & 0 & 1 & 0 & 0 \\
	0 & \frac{s}{\rho _{0[2]}} & 0 & 1 & 0 & 0 \\
	0 & 0 & \frac{s}{\rho _{0[3]}} & 1 & 0 & 0 \\
	-\frac{C_{T [1]}^2 q^2}{\rho _0} & -\frac{C_{T [2]}^2 q^2}{\rho _0} & -\frac{C_{T [3]}^2 q^2}{\rho _0} & s & -q^2\left(\frac{C_{T [1]}^2 \rho _{0[1]}}{T_0
		\rho _0}+\frac{C_{T [2]}^2 \rho _{0[2]}}{T_0 \rho _0}+\frac{C_{T [3]}^2 \rho _{0[3]}}{T_0 \rho _0}\right) & -q^2 \\
	0 & 0 & 0 & \frac{2}{3} & \frac{s}{T_0} & 0 \\
	4 G \pi  & 4 G \pi  & 4 G \pi  & 0 & 0 & q^2 \\
	\end{array}
	\right).
$$
%\end{center}

\noindent This matrix leads to the dispertion relation:
\begin{equation}
s^3[s^2+q^2\left(C_{T [1]}^2\frac{\rho _{0[1]}}{\rho _0}+\frac{5}{3} C_{T[2]}^2\frac{\rho _{0[2]}}{\rho _0}+\frac{5}{3} c_{T[3]}^2 \frac{\rho _{0[3]}}{\rho_0}\right)-4 \pi G \rho _0]=0
\end{equation}

\noindent where $\rho_0=\rho_{0[1]}+\rho _{0[2]}+\rho_{0[3]}$. The corresponding Jeans wave number reads:
\begin{equation}
q^2= \frac{ 4 \pi  G \rho _0}{\frac{5}{3}\left(C_{T [1]}^2 \frac{\rho _{0[1]}}{\rho _0}+ C_{T[2]}^2\frac{\rho _{0[2]}}{\rho _0}+C_{T[3]}^2 \frac{\rho _{0[3]}}{\rho_0}\right)}
\end{equation}
In general, for a multicomponent dilute inert mixture of $n$ components, the critical wavenumber $q_{c}$ reads:
\begin{equation} 
q^2_c= \frac{ 4 \pi  G \rho_0}{\frac{5}{3}\sum _{i=1}^n
	\left(\frac{\rho _{0[i]}}{\rho _0}C_{T [i]}^2\right)},\qquad \rho_0=\sum_{i=1}^n \rho_{0[i]}.
\end{equation}
Although similar results have been obtained since the pioneeering work of Grishchuk and Zel’dovich  \cite{GZ}, there are quite few works in which kinetic theory has been directly applied to the analysis of the gravitational instability of dilute mixtures of gases \cite{GMK}. The approach here presented leads to remarkably simple expressions for the eigenfrequencies of the dynamical system and for the critical wave number. This type of representation  for the transport equations motivates further analysis of the Jeans instability  involving difusssion and other dissipative effects in different temperature scales. This will be the subject of future work.

\section{Acknowledgements}
The authors wish to thank J.H. Mondragón-Suárez for his valuable comments to this manuscript. \\
This work has been supported by the Applied Research Institute of Technology
(INIAT) of U. Iberoamericana, Mexico.

\end{document}